\documentclass[epj]{webofc}
\usepackage[varg]{txfonts}   % Web of Conferences font
\usepackage{hyperref}
\usepackage{gensymb}

\renewcommand*{\eqref}[1]{Eq.~(\ref{eq:#1})}

\newcommand*{\figref}[1]{Fig.~(\ref{fig:#1})}
\newcommand*{\figlab}[1]{\label{fig:#1}}

\newcommand*{\seclab}[1]{\label{sec:#1}}
\wocname{\includegraphics[width=0.25cm,clip]{logoARENA} ARENA2018}
\wocname{ARENA2018}
\woctitle{ARENA2018}
\woctitle{\includegraphics[width=0.25cm,clip]{logoARENA} ARENA2018}
\begin{document}

\title{\vskip-8em
Towards online triggering for the radio detection of air showers using deep neural networks
}
%
% subtitle is optional
%
%%%\subtitle{Do you have a subtitle?\\ If so, write it here}

\author{
\firstname{Florian} \lastname{F\"uhrer}\inst{1,2}\fnsep\thanks{\email{fuhrer@iap.fr}}
\and  \firstname{Tom} \lastname{Charnock}\inst{1}
\and \firstname{Anne} \lastname{Zilles}\inst{1}
\and \firstname{Matias} \lastname{Tueros}\inst{3}
}
%%%%%%%%%%%%%
\institute{Sorbonne Universit\'{e}, UPMC Univ.  Paris 6 et CNRS, UMR 7095, Institut d'Astrophysique de Paris, 98 bis bd Arago, 75014 Paris, France
\and
           Sorbonne Universit\'e, Institut Lagrange
de Paris, 98bis boulevard Arago, 75014 Paris, France
\and
           Instituto de F\'{i}sica La Plata - CONICET/CCT- La Plata. Calle 49 esq 115. La Plata, Buenos Aires, Argentina
          }

\abstract{%
The detection of air-shower events via radio signals requires the development of a trigger algorithm for clean discrimination between signal and background events in order to reduce the data stream coming from false triggers.
In this contribution we will describe an approach to trigger air-shower events on a single-antenna level as well as performing an online reconstruction of the shower parameters using neural networks.}
\maketitle
\section{Introduction}
\label{intro}
Machine learning and, in particular, deep learning using neural networks (NN)
is a field which is currently beginning to thrive in
astronomy.
This means we can make use of well-established techniques to tackle the challenge of both self-triggering on radio signals, preferably on the single antenna level, and performing online parameter-reconstruction of neutrino and ultra high energy cosmic ray (UHECR) induced air-shower events based on detected signals in an array of antennas.
There is expected to be a large amount of human-made background noise (mostly transient noise) in the radio-data which needs to be dismissed efficiently to distinguish between signals from air showers induced by cosmic primaries and neutrino events.
As an example of the scale of the problem, we expect $\sim1$ neutrino event-per-year for one of the proposed
$10\,000$ antennas sub-arrays of the GRAND experiment~\cite{GRAND}, while the trigger rate due to background will be $\sim$kHz.
%Current experiments work with a trigger threshold to identify signals which is defined by the ratio of the signal strength to the noise level. Here, to be detectable, the signal has to be several times larger than the standard deviation of the noise distribution, which is dominated by galactic and antenna noise.
%

%Through using machine learning we propose that it would be feasible to lower the detection threshold for a possible signal significantly by efficient learning of the pulse shape of an expected signal and a good discrimination of transient noise.
Through the use of machine learning we propose that the false event detection rate can be reduced significantly, which reduces the data-stream rate and effectively lowers the triggering threshold.
Lowering the triggering threshold would allow less-energetic air showers to be detected which would lead to an important gain in statistics.
The feasibility of self-triggering on air-shower events via the
measurement of the radio signals is reinforced by the reports of
\cite{TREND,Auger}.
The purpose of this proceeding is to demonstrate the feasibility of NN based
triggering and to motivate further research in this direction.
%Here we will highlight the advantages of NN based triggering compared to
% conventional trigger algorithms.
%

\section{Neural networks as triggers}
\seclab{strat}
\begin{figure}
\centering
%\vskip-1em
\includegraphics[width=0.7\textwidth]{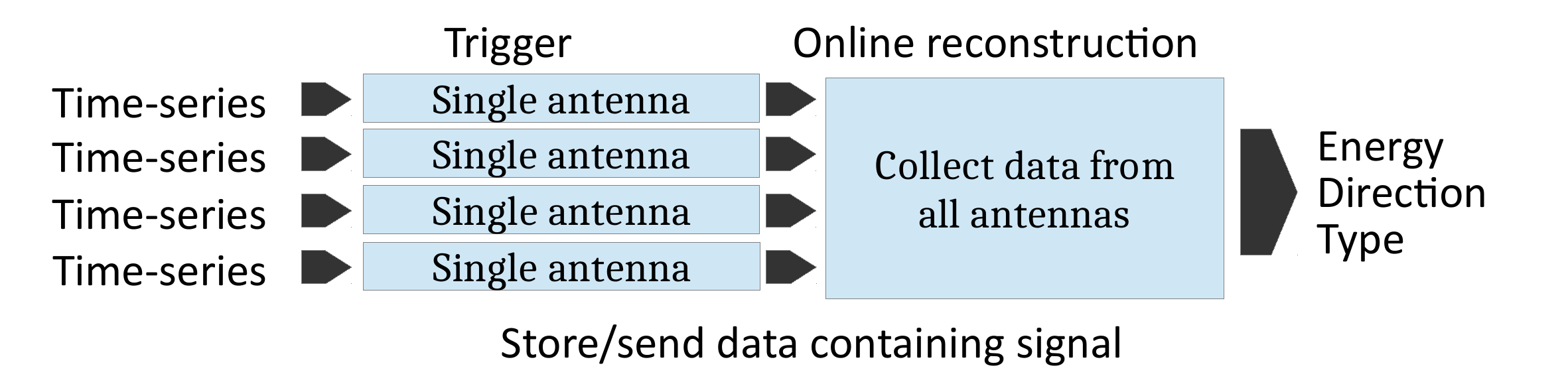}
\caption{A sketch of the considered pipeline of online triggering and
reconstruction.
At each antenna a NN searches the incoming data stream
for a radio signal.
The discovered signals are passed together with a time stamp to a second NN, which can then perform the online reconstruction of the shower
parameters.}
\figlab{stratSketch}
\vskip-2em
\end{figure}

We assume the simplified two step data taking process of a radio array depicted
in \figref{stratSketch}. The process starts with a trigger on a single antenna
level, and ends by the reconstruction of the shower parameters. In this proceeding we will describe a NN trigger which searches the data recorded from the antenna in a given time interval for a signal. In a future publication we will describe the reconstruction network which takes the collected time traces containing signals from all antennas as an input and reconstructs the primary particle's type and parameters.

\subsection{Deep learning}
\seclab{DL}

In machine learning, ``deep learning'' allows for abstraction between the output
and the provided input data where the non-trivial function mapping the input to the output is fitted \cite{Goodfellow2016}.
Since the neural network learns directly from the data that is provided, the only necessary products are simulations of the real data.
Provided with such simulations we expect a NN based trigger
to better generalise than other methods, such as matched filters that
necessitate templates for each event type. %In contrast to other triggering methods, such as
% matched filters, we expect a NN based trigger to better generalise because, as long as simulations of realisitic events are available, the network will be able to learn about the data, whereas templates for the matched filters will be necessary for each individual event type.
As an additional advantage, a NN based trigger will have a lower computational cost than using matched filters \cite{George2017}.
\\

\noindent
\textbf{Training data:} To demonstrate
the feasibility of online triggering using NN, we consider a GRAND-like toy array consisting of $35 \times 35$ antennas on slope of $5 \degree$.
Our simulated signals consist of three independent voltage trace components which are folded with the antenna response and Gaussian white noise with a standard deviation of $15\: \mu V$ and then filtered to $50-200\:\text{MHz}$.
For the UHECR we use protons and iron-nuclei with an energy of $0.1-100\:\text{EeV}$
and a zenith angle of $65 - 85 \degree$.
The neutrino distribution in direction and energy follow the expected distribution for the GRAND array.
\\

\noindent
\textbf{Some notes on the generalisation to real data:}
A NN trained on white noise is likely to have a similar performance on realistic noise if the network can adapt to the realistic noise.
Thanks to transfer learning, where a network fully trained on simulations is essentially tuned by training with a small amount of real data, the network adaptation is possible without retraining from scratch \cite{Transfer}.
Given that the network can learn about realistic noise, there is still the obstacle of transient noise closely mimicking real signals.
This does not pose a problem as long as the transient noise can be measured and
hence learned, allowing the network to distinguish between the minute
differences between different samples of data.
Given that we do not expect most transient signals to exactly mimic true signals
we are confident that the network will be able to filter most transient
noise\footnote{We expect to profit from the fact that the use of
polarization information is already built into the NN architecture.}.

\section{Results}
\seclab{results}
For the trigger, we trained a convolutional NN with three convolutional layers and one fully connected hidden layer with 64 units.
The convolutional layers have consecutive kernels sizes of 16, 8 and 8 and 16, 32 and 64 filters and we perform a max-pooling over consecutive time elements.
We train the network by minimising the categorical cross entropy.
The neural network is written using TensorFlow for its inter-platform and co-GPU
and -CPU portability.% and the training is performed on an NVIDIA Quadro P6000.
%The training itself takes approximately *** to complete.

The NN is trained using $\sim 150\,000$ time traces half containing a signal and half containing only noise.
We found that the number of correctly classified traces using the NN is 72\%, which is higher than when using a threshold trigger at $60\:\mu V$ applied to each of the three voltages traces independently, where the accuracy is 69\%.
While the true positive classification is only slightly improved by the NN, 43\% compared to 42\%, the number of false triggers are reduced by more than one order of magnitude, from 4\% to less than 0.2\%.
Since we wish to reduce the data-stream efficiently it is exactly the rate of
false events that we want to reduce. %, even if we do not improve greatly on the
% number of events that are detectable.
A detailed comparison of the two trigger algorithms is beyond the scope of this
proceeding, we postpone this together with an accurate calculation of the
expected data stream to a future publication.

\section{Conclusion and outlook}
\seclab{conclusion}
In this proceeding we have described the first steps towards self-trigger on
radio signals using a NN.
This NN-based trigger performs better than a conventional threshold trigger, in
particular reducing the number of false alarms drastically and therefore
potentially making the data-stream of signal detections more pure. Our promising results motivate further exploration of the existing potential to improve the NN trigger.
Of course the NN can be taught to better generalise through the gathering of more data or perform better via hyperparameter tuning, but another possibility is to split the trigger into two levels.
The first level of a two level trigger algorithm would be the NN described in this proceeding and the second level would use the coincidences in the arrival time among nearby antennas to further reduce the false event rate.
However, if this second level trigger is built into the online reconstruction of
the particle shower using the an entire GRAND sub-array, the accidental
over-triggering due to noise can essentially be absorbed into the reconstruction network (mentioned earlier) which is inherently more resilient to artefacts due to the way it correlates all antennas.
A complementary, and perhaps more simple, modification to the level-one trigger
could be to train the NN specifically to do online denoising of voltage
traces directly from the observed traces.

Even though the proposed network seems much more computationally heavy than a simple threshold trigger, advancements in low power, highly optimised machine learning chipsets actually suggests that there will be little to no difference in computational complexity hence making the process of NN-based triggering scalable.

\section*{Acknowledgement}
%This work has been done within the Labex ILP (reference ANR-10-LABX-63) part of
%the Idex SUPER, and received financial state aid managed by the ANR, as part of
%the programme Investissements d'avenir under the reference ANR-11-IDEX-0004-02
% .
This work has received support from the French Agence Nationale de la
Recherche under the references ANR-10-LABX-63,
ANR-11-IDEX-0004-02 and ANR-16-CE31-0001. This work has made use of the Horizon
Cluster hosted by the Institut d'Astrophysique de Paris. The Quadro P6000 used in this study was donated by the
NVIDIA Corporation.
%
% BibTeX or Biber users please use (the style is already called in the class, ensure that the "woc.bst" style is in your local directory)
% \bibliography{name or your bibliography database}
%
% Non-BibTeX users please use
%

\end{document}